\documentclass[aps,prb,reprint,groupedaddress]{revtex4-1}

\usepackage[T1]{fontenc}
\usepackage[utf8]{inputenc}
\usepackage{graphicx}
\usepackage{bm}
\usepackage{comment}
\usepackage{makecell}
\usepackage{url}
\usepackage{hyperref}
\usepackage{natbib}
\usepackage{setspace}

\begin{document}
 \begin{spacing}{1}

\title{Homogeneous and inhomogeneous sources of optical transition broadening in room temperature CdSe/ZnS nanocrystal quantum dots }

\author{M. Wolf, J. Berezovsky}
\email[Corresponding author:]{jab298@case.edu}
\affiliation{Department of Physics, Case Western Reserve University, Cleveland, Ohio 44106}
\date{\today}
\begin{abstract}
We perform photoluminescence excitation measurements on individual CdSe/ZnS nanocrystal quantum dots (NCQDs) at room temperature to study optical transition energies and broadening.  The observed features in the spectra are identified and compared to calculated transition energies using an effective mass model.  The observed broadening is attributed to phonon broadening, spectral diffusion and size and shape inhomogeneity.  The former two contribute the broadening transitions in individual QDs while the latter contributes to the QD-to-QD variation.   We find that phonon broadening is often not the dominant contribution to transition line widths, even at room temperature, and that broadening does not necessarily increase with transition energy.  This may be explained by differing magnitude of spectral diffusion for different quantum-confined states.
\end{abstract}
\maketitle

Optical absorption in room temperature colloidal nanocrystal quantum dots (NCQDs) is of key importance for applications of these materials including photovoltaics,\cite{klimov2004,kamat2008}  light emitting diodes,\cite{wood2010} and fluorescent tagging.\cite{Bruchez1998, Michalet2005}  A particularly useful feature of these materials is that the optical transitions are broadened over a range comparable to the spacing between them, such that they strongly absorb light over a wide energy range.  This feature, however, makes it difficult to study the origins of the broadening because individual transitions are difficult to discern.  Previous work has characterized the transitions for ensembles of NCQDs of various radii. \cite{norris1996} Here, we investigate the spectrum of transitions in individual nanocrystals, using a photoluminescence excitation (PLE) technique.\cite{oliveria1995, norris1996}  The experiment reveals the broadening of the transitions in the absence of ensemble averaging, and also the variation in transition energies that gives rise to the ensemble broadening.  We identify the observed transitions by comparison to calculated transition energies using an effective mass calculation. The results can be explained by three mechanisms: phonon broadening \cite{goupalov2001}, spectral diffusion \cite{neuhauser2000}, and size inhomogeneity.\cite{norris1996}  The strength of latter two effects both depend on the sensitivity of the transition energy to changes in the confining potential. In general, this sensitivity increases with increasing transition energy, resulting in increased broadening.  However, we see evidence that some transitions do not follow this trend -- a fact explained by mixing of the different valence sub-bands in the valence band eigenstates. 

\begin{figure}[htbp]
	\centering
	\includegraphics[width=0.44\textwidth]{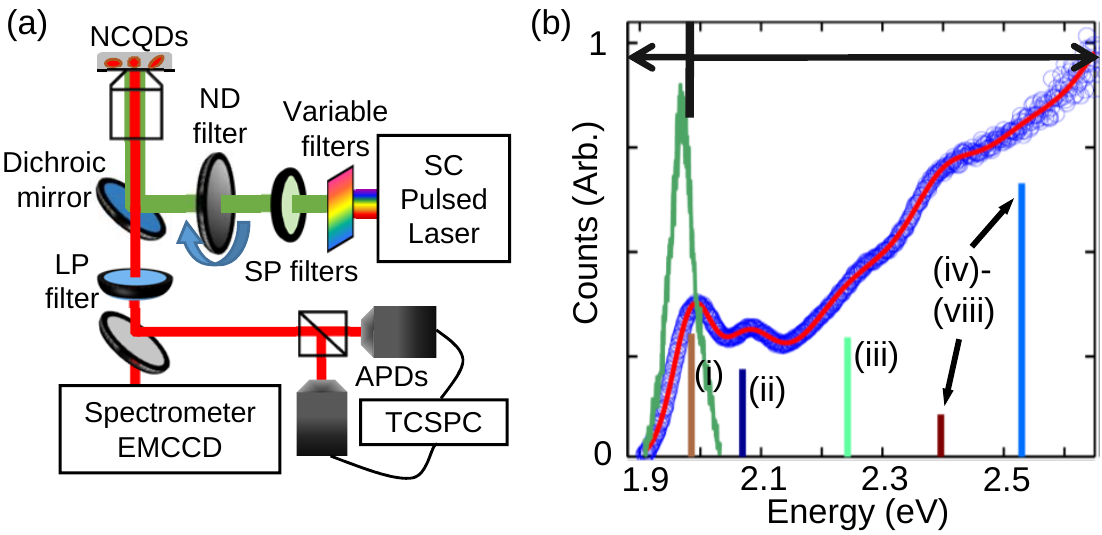}
	\caption{(a)Schematic of experiment set-up. (b) Absorption spectrum of a solution of QDs (blue) with fit of  Eq.~\ref{eq:fitfn} (red) and NCQD's PL spectrum (green).  The vertical black line corresponds to the LP cutoff filters.  The left region of the black line are the emission energies collected by the detectors while the right region are the excitation energies being swept.  The bottom vertical lines pertain to the energy transitions in Fig.$\ref{fig3}$ (a) with the height of the bar indicating oscillator strength.}
	\label{fig1}
\end{figure}

Figure~\ref{fig1} shows a schematic of the experimental setup.  Excitation is provided by a supercontinuum fiber laser (Fianium SC450PP), outputting $\sim 25$~ps duration pulses at a repetition rate of 5~MHz.  The spectrally broad output of the laser is then filtered by a linearly-graded high-pass and low-pass filter mounted on motorized translation stages to tune the cut-on and cut-off wavelengths.  The excitiation wavelengths range from 465 to 610 nm with a bandwidth of 5nm (FWHM). The excitation beam is passed through two short pass (SP) filters with cut-off wavelength of 610 and 615 nm.  A constant excitation power $P \cong 1.37 \mu$W (intensity $I \sim 500$~W/cm$^2$) was maintained by a continuously-graded neutral density (ND) filter wheel on a motorized rotation stage, calibrated over the entire wavelength range.  The excitation light is focused onto the sample through an oil immersion objective lens.  PL from the sample, centered around 630 nm, is collected via the same microscope objective, and passed through two long pass (LP) filters.
 
\begin{figure}[htbp]
	\centering
	\includegraphics[width=0.47\textwidth]{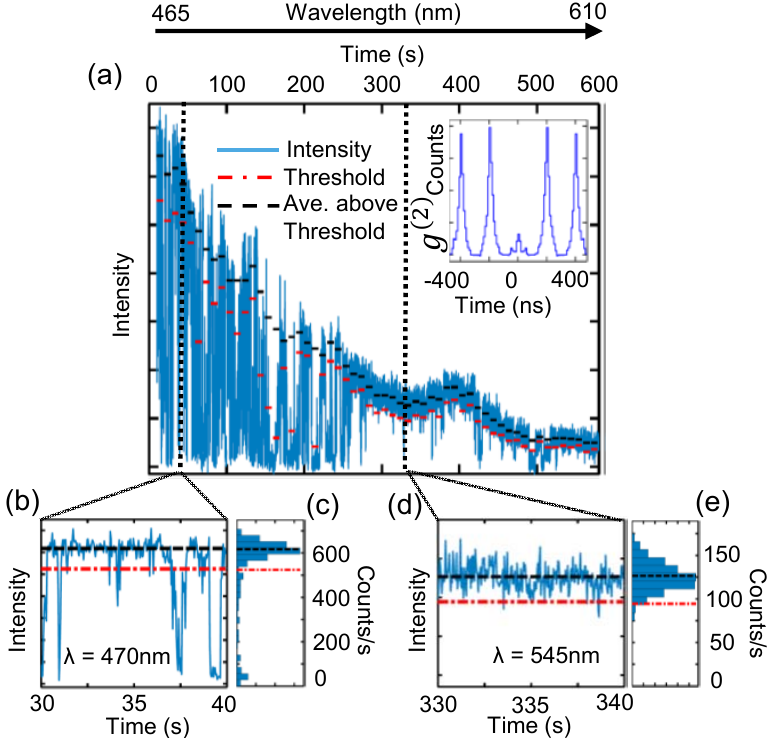}
	\caption{(a) Typical blinking trace (solid) for single QD with the threshold and the average counts above the threshold.  The inset shows the cross-correlation. (b) and (d) show expanded view of two PL traces at different wavelengths. (c) and (e) are histograms of the PL traces indicating the threshold (dashed dotted) and average above threshold (dashed).}
	\label{fig2}
	\end{figure}

The PL is imaged on an electron-multiplication CCD (EMCCD) camera, or sent to a pair of avalanche photodiodes (PDM-50ct) collected by time-correlated single photon counting system (TCSPC, Hydraharp 400).  The TCSPC allows for both a 2-photon cross-correlation $g^{(2)}$ measurement (Fig.~\ref{fig2}(a) inset) and an intermittency (blinking) trace.\cite{nirmal1996, janko2013, galland2011}

The PLE measurement is carried out by sweeping the excitation energy, and at each value of the energy, collecting photon counts for 10 s.  Figure~\ref{fig1}(b) illustrates the energy ranges involves in the experiment, in relation to the PL spectrum (green) and the ensemble absorption (blue).  Emission is collected at energy below the vertical black line, and excitation energy is swept at energies above the vertical black line.  A trace of PL vs.~time is then constructed from the measured photon counts by binning photon arrival times as shown in Fig.~\ref{fig2}(a).  The PL trace exhibits blinking as the count rate transitions between a high level and near zero, as is typical in NCQDs \cite{nirmal1996}.  Figures~\ref{fig2}(b) and (d) show an expanded view of the PL trace at two different excitation wavelengths.  In the first, blinking is clearly evident.  In the second, the QD is in the high emission state the entire time -- it does not blink off.  This is seen by constructing a histogram of count rates, as shown in Fig.~\ref{fig2}(c) and (e).  In Fig.~\ref{fig2}(c) the on and off states appear as a peak around 620 counts/s, and near zero, respectively.  In  Fig.~\ref{fig2}(e) there is a peak in the histogram around 125 counts/s, and no peak near zero.  Measurements without blinking are common at lower excitation energy, due to the dependence of blinking statistics on excitation rate and energy \cite{crouch2010}.  To extract the emission rate from the high level only, we exclude the time the NCQD spends in the low-emission state.  To do this, we set a threshold for each excitation energy, and only count the emission rate above this level. This threshold is determined from the count rate histograms (e.g. Fig.~\ref{fig2} (c) and (e)), by finding the maximum of the upper peak and setting the cutoff a specified number of bins lower.  Then by taking the average counts above the threshold, a single-QD PLE spectrum is created.  Some QDs occasionally transition into a state with emission intermediate between the high and low levels.  If this occurs for a single excitation energy point, this point is removed from the spectrum and replaced by the linear interpolation of its neighbors. We do not include QDs with two or more consecutive ambiguous points in the dataset.

The NCQDs studied in this experiment are CdSe/ZnS core/shell colloidal quantum dots (NN-Labs CZ620-10).  The core has a typical radius of 3 nm, overcoated by a 1.5 nm shell.  The core diameter varies by approximately 10\% across the ensemble.  The optical absorption spectrum of the QDs in solution is shown in Fig.~\ref{fig1} (b), with the lowest absorption peak occurring at about 2 eV.  The QDs are diluted in toluene with PMMA, and spin-coated onto a glass coverslip.  The dilution of QDs is chosen to produce a concentration less than 1 QD per (10 $\mu$m$)^{2}$, to optically address individual QDs.  A high-concentration sample is prepared in the same way, so that many QDs are probed within the laser spot, for a reference ensemble measurement.

In order to understand the transitions that give rise to the observed absorption and PLE spectra, we calculate electron and hole wavefunctions and energies in a spherical effective mass model.  The calculations follow Refs.~\onlinecite{ekimov1993, chepic1990, efros2000}, with the exception that we include the shell layer for the conduction band states.  The hole states are composed of mixed states of the heavy, light, and split-off hole valence  bands, whereas the electron states arise from the conduction band.  For the electron states, we take the confining potential to be spherically symmetric, with a 0.9 eV step at the core/shell boundary, and an infinite barrier at the outer shell boundary.\cite{Dabbousi1997}  For the hole states, we assume that the core/shell interface represents an infinitely high barrier, due to the larger effective mass of the holes as compared to the electron.  We neglect the effects of electron-hole exchange interaction, crystal field splitting, and shape anisotropy.  These effects shift energy levels by up to several tens of meV, which sets a bound on the precision of the energies calculated here. \cite{efros2000} The oscillator strengths and energies of the interband transitions are then calculated, with the Coulomb energy included empirically as in Ref. \onlinecite{norris1996}.
 
\begin{figure}[htbp]
\centering
	\includegraphics[width=0.43\textwidth]{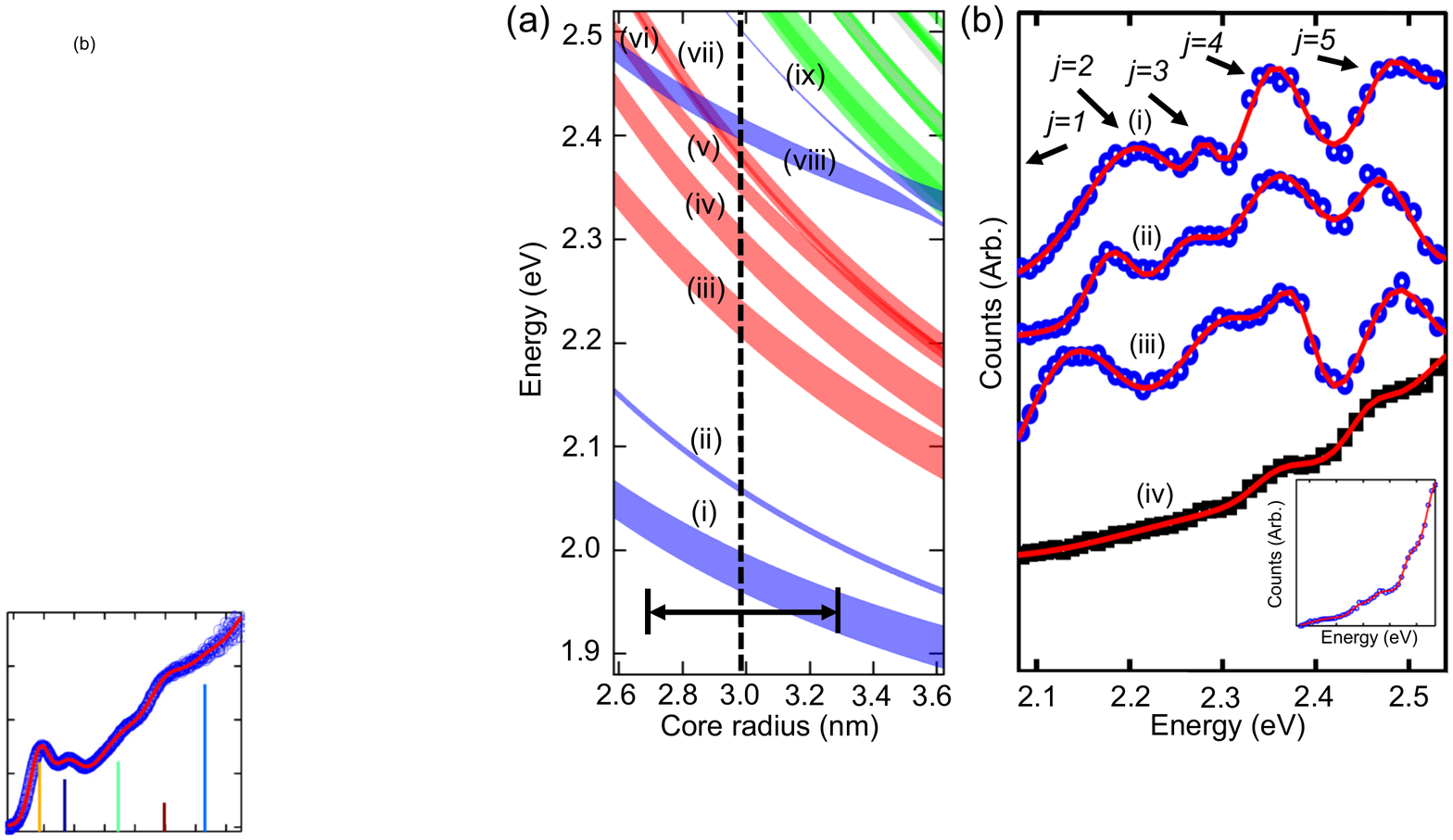} 
	\caption{(a) Calculated energy transitions vs.~core radius. The approximate mean QD radius is indicated by vertical dashed line with the standard deviation given by the double arrow.  Transition assignments:(i) 1S$_e$-1S$_{3/2}$, (ii) 1S$_e$-2S$_{3/2}$, (iii) 1P$_e$-1P$_{3/2}$, (iv) 1P$_e$-1P$_{5/2}$, (v) 1P$_e$-1P$_{1/2}$, (vi) 1P$_e$-2P$_{3/2}$, (vii) 1P$_e$-2P$_{5/2}$, (viii) 1S$_e$-2S$_{1/2}$, (ix) 1S$_e$-3S$_{1/2}$. (b) (i)-(iii) Single QD PLEs with corresponding fits. (iv) Ensemble PLE with fit.  Inset shows a single QD PLE with no distinct features.}
	\label{fig3}
	\end{figure}

Fig. \ref{fig3} (a) displays the calculated transition energies vs. core radius, with the oscillator strength represented by the thickness of the line.  Transition notation follows Refs. \onlinecite{norris1996,efros2000}.  The dashed black line in Fig.~\ref{fig3} (a) indicates the mean core size in the present measurements, with the horizontal double arrow showing the approximate standard deviation of core sizes in the sample.

To confirm that the calculated transition energies match the relevant transitions in our sample, we fit the measured absorption spectrum to a sum of Gaussian peaks:
\begin{equation}
\alpha(E) = \sum_{j=1}^{N} a_j \exp{\left( \frac{-(E-E_j)^2}{2\Gamma_j^2}\right)}+f_{bkg}(E).
\label{eq:fitfn}
\end{equation}

Here, $N=5$ and $a_j$, $E_j$, and $\Gamma_j$ are varied in the fit, and the $E_j$s are constrained within bounds set by the calculated transition energies.  To account for additional weak transitions not captured by this simple theoretical treatment, a broad Lorentzian background term $f_{bkg}$ is also included in the fit function.  The resulting fit, Fig.~\ref{fig1} (b), matches the data well, with transition energies in agreement with the calculated values.  We assign the three lowest energy transitions to the 1S$_e$-1S$_{3/2}$, 1S$_e$-2S$_{3/2}$, and 1P$_e$-1P$_{3/2}$ transitions.    Above those three transitions, we expect several closely spaced transitions which may not be distinguishable in the ensemble measurement.  The fourth and fifth peaks in Eq.~\ref{eq:fitfn} are thus assigned to some combination of the 1P$_e$-1P$_{1/2}$, 1P$_e$-1P$_{5/2}$, 1P$_e$-2P$_{3/2}$, 1P$_e$-1P$_{5/2}$, and 1S$_e$-2S$_{1/2}$ transitions.

We use a fitting procedure for the single-QD PLE data similar to that for the ensemble absorption data. The PLE spectra covers an energy range from 2.1 to 2.55 eV, which does not include the two lowest transitions ($E_1$ and $E_2$) because near-resonant excitation overwhelms the detectors. We fit to Eq.~\ref{eq:fitfn} with $N=5$, where the $j=1$ peak is outside the measured energy range and captures the high-energy tails of the 1S$_e$-1S$_{3/2}$, 1S$_e$-2S$_{3/2}$ transitions. To ensure $E_j$ values that increase monotonically, the $E_j$s are constrained within increasing energy windows which overlap by several meV. From these fits, we obtain peak energies $E_j$, widths $\Gamma_j$, and associated 95\% confidence intervals $\delta E_j$ and $\delta \Gamma_j$.  Fig.~\ref{fig3} (b) shows four characteristic PLE traces (blue) obtained on different QDs with the associated fits and an ensemble PLE (black). In all cases, the fit function describes the data well. Spectra (i)-(iii) show four peaks, $j=2-5$, while the inset shows a spectrum with less prominent peaks. The peak energies $E_j$ and widths $\Gamma_j$ in fits (i)-(iii) are determined with $\delta E_j/E_j<0.05$ and $\delta \Gamma_j/\Gamma_j <0.5$, whereas the inset fit fails to determine these quantities within any reasonable confidence interval. In total, we have measured PLE spectra on 93 QDs. Eq.~\ref{eq:fitfn} provides visually good fits to all the data, with about 22 of those providing reliable fit parameters for energy and width as in fits (i)-(iii), as judged by the confidence intervals.  Spectrum (iv) shows an ensemble PLE spectrum measured in the same setup, using the same technique. In comparison, the ensemble has no prominent features with the fit being extremely overdetermined.
%

By comparison of the fit parameters with the calculated transition energies, we can assign the observed structure of the PLE spectra to optical transitions. The $j=2$ and $j=3$ peak are likely to arise from the (iii) 1P$_e$-1P$_{3/2}$ and (iv) 1P$_e$-1P$_{5/2}$ transition respectively, while the $j=5$ peak is attributed to the (vii) 1S$_e$-2S$_{1/2}$ transition. The $j=4$ feature likely arises from a combination of the (v) 1P$_e$-1P$_{1/2}$, (vi) 1P$_e$-2P$_{3/2}$, and (vii) 1P$_e$-2P$_{5/2}$ transitions. The lack of clear peaks in some of the PLE spectra can be explained by QDs with size in the range where the 1S$_e$-2S$_{1/2}$ transition overlaps with some of the transitions to the 1P$_e$ state, Fig. \ref{fig3} (a), or in QDs with particularly large deviation from spherical symmetry.


\begin{figure}[htbp]
	\centering
	\includegraphics[width=0.28\textwidth]{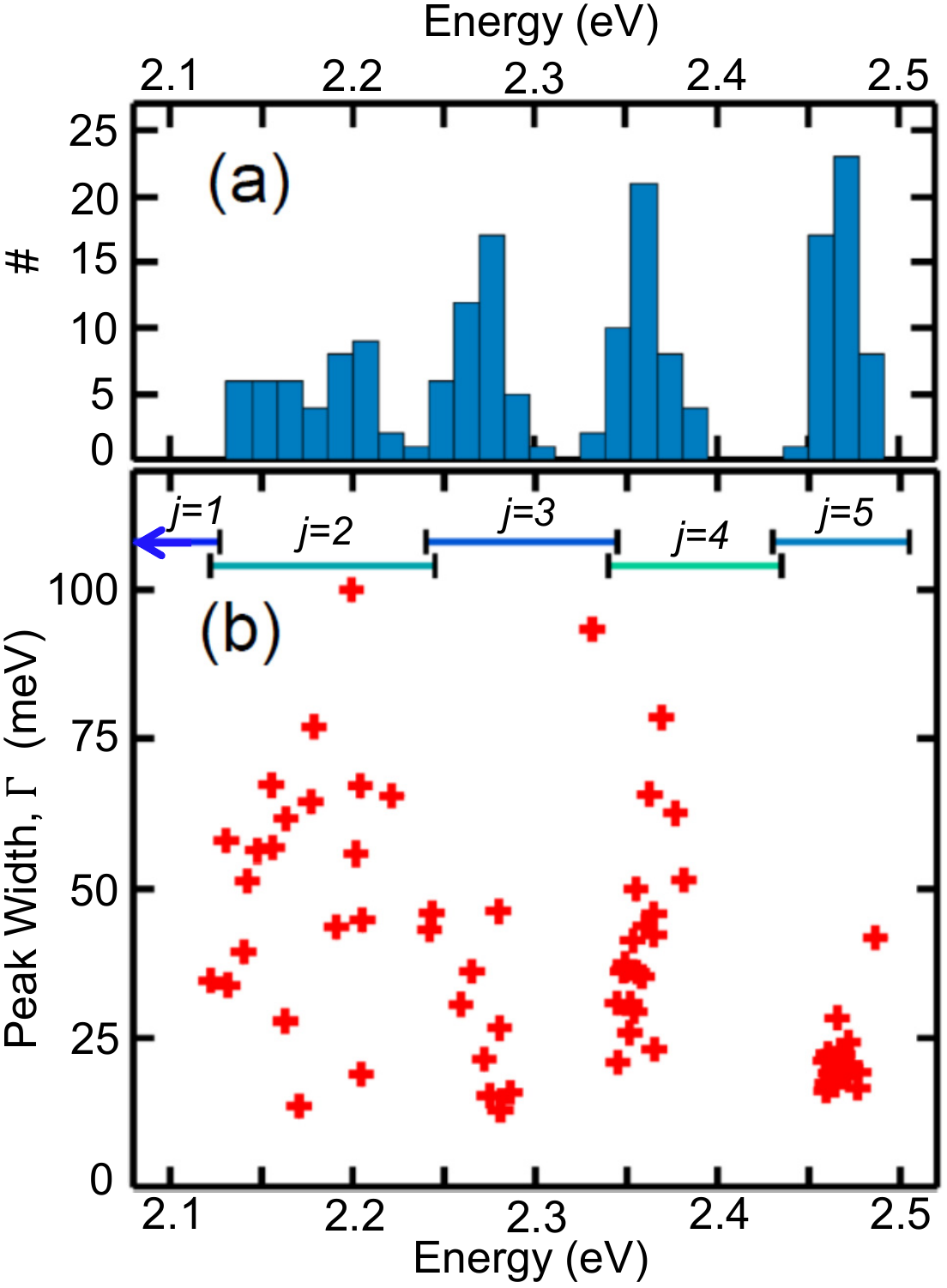}
	\caption{(a) The transition energies from fits with error less $5\%$. (b) The transition energy versus the peak width from fits with energy and width error $\delta E_j/E_j<0.05$ and $\delta \Gamma_j/\Gamma_j<0.5$.}
	\label{fig5}
	\end{figure}  

Fig.~\ref{fig5}(a) shows a histogram of $E_j$ values for all peak energies with $\delta E_j/E_j<0.05$ containing 177 total values.  Four distinct clusters are visible, corresponding to $j=2-5$.   The lower energy peak, $E_{2}$, has the broadest range of $\approx 110$ meV indicating significant QD-to-QD variation, while the highest energy peak, $E_{5}$, has a range of $\approx 50$ meV, indicating less QD-to-QD variation.  The separation between the clusters becomes more pronounced with increasing energy.  The separation between $E_4$ and $E_5$ has a clear gap of about 100 meV, while the separation between $E_2 $ and $E_3$ is less defined. The clear separation between the clusters may partly be an artifact due to the removal (by the confidence interval criterion) of NCQDs that do not show four clear peaks.  Nevertheless, the differences in the energy range covered by different clusters may give some indication of the inhomogeneity of transition energies within the ensemble.  

To study the distribution of peak widths $\Gamma_j$ and their correlation with energies $E_j$, Fig.~\ref{fig5}(b) shows a scatter plot of $\Gamma_j$ vs. $E_j$. Here we show only points with $\delta E_j/E_j<0.05$ and $\delta \Gamma_j/\Gamma_j<0.5$ consisting of 68 points. $\Gamma_5$, corresponding to the highest energy transition ($j=5$) has the smallest variation and smallest average $\langle\Gamma_5 \rangle = 21$~meV.  The $j=2$ and $j=4$ peaks have a significantly larger variations of widths spanning the range from 20-100 meV.  The $j=3$ width variation is about 15 - 45 meV.  

We attribute the observed transition broadening and QD-to-QD variability to three effects: 1. phonon broadening, 2. spectral diffusion, and 3. size and shape inhomogeneity.  The first two are present in individual QDs, and hence contribute to the broadening, $\Gamma$, seen in the PLE spectra.  The third is an ensemble averaging effect, and thus contributes to QD-to-QD variability of $E_{j}$, which manifests as an additional source of broadening in ensemble measurements. 

We can estimate the magnitude of the different contributions to transition broadening.  First, phonon broadening $\sim kT$, which sets a minimum linewidth. \cite{goupalov2001} Second, the inhomogeneous broadening depends on the sensitivity $s_j= |dE_j/da|$ of each transition energy $E_j$ to QD radius $a$.  The quantity $s_{j}$ can be seen as the slope of the different transitions in Fig. \ref{fig3}(a).  With some distribution of $a$ across the ensemble, transitions with higher $s_j$ will show greater inhomogeneous broadening (or QD-to-QD variation).  In the transitions that give rise to the PLE spectra measured here, the transitions (iii)-(vii) have significantly higher $s$ than transition (viii).  This occurs mainly because transitions (iii)-(vii), shown in red in Fig.~\ref{fig3}, involve the $1\mathrm{P_e}$ conduction band level whose energy depends more strongly on confinement than the $1\mathrm{S_e}$ level involved in transition (viii).  The higher energy of transition (viii), despite involving the lowest energy conduction band level, occurs because the valence band state involved in this transition ($2\mathrm{S_{1/2}}$) mainly arises from the split-off valence band.  Third, spectral diffusion arises from random rearrangements of the local charge distribution in the environment of a QD,\cite{empedocles1997} resulting in energy shifts.  Spectral diffusion of the lowest energy transition, measured via the PL energy, has been observed with a magnitude $\sim 10$~meV.\cite{emp1999, neuhauser2000}  It is possible that the magnitude of spectral diffusion for higher energy transitions may depend on the parameter $s_j$ as a measure of the sensitivity of the energy levels to changes in the confining potential.  If so, this could explain the larger observed broadenings $\Gamma_j$ for $j=2-4$, corresponding to transitions (iii)-(vii).  Likewise, the smaller values of $\Gamma_5$, and the smaller range of $E_5$, are consistent with the smaller value of $s_5$, from transition (viii), $2\mathrm{S_{1/2}1S_e}$.

We used a photoluminescence excitation technique to probe the absorption spectra of individual quantum dots at room temperature.  We observe additional peaks in the single-QD spectra not visible in the ensemble absorption spectrum.  By measuring the energies and widths of these peaks, we reveal information about broadening mechanisms of transitions in these structures.  We find that some transitions are broadened significantly more than $kT$, likely due to spectral diffusion.  We also find that transition broadening does not necessarily increase with energy, a fact possibly explained by differing spectral diffusion amplitude for different quantum-confined energy levels.
   
\begin{acknowledgments}
This work was supported by DOE, award No. DE-SC0008148.
\end{acknowledgments}
\end{spacing} 
\bibliography{References}

\begin{thebibliography}{19}%
\makeatletter
\providecommand \@ifxundefined [1]{%
 \@ifx{#1\undefined}
}%
\providecommand \@ifnum [1]{%
 \ifnum #1\expandafter \@firstoftwo
 \else \expandafter \@secondoftwo
 \fi
}%
\providecommand \@ifx [1]{%
 \ifx #1\expandafter \@firstoftwo
 \else \expandafter \@secondoftwo
 \fi
}%
\providecommand \natexlab [1]{#1}%
\providecommand \enquote  [1]{``#1''}%
\providecommand \bibnamefont  [1]{#1}%
\providecommand \bibfnamefont [1]{#1}%
\providecommand \citenamefont [1]{#1}%
\providecommand \href@noop [0]{\@secondoftwo}%
\providecommand \href [0]{\begingroup \@sanitize@url \@href}%
\providecommand \@href[1]{\@@startlink{#1}\@@href}%
\providecommand \@@href[1]{\endgroup#1\@@endlink}%
\providecommand \@sanitize@url [0]{\catcode `\\12\catcode `\$12\catcode
  `\&12\catcode `\#12\catcode `\^12\catcode `\_12\catcode `\%12\relax}%
\providecommand \@@startlink[1]{}%
\providecommand \@@endlink[0]{}%
\providecommand \url  [0]{\begingroup\@sanitize@url \@url }%
\providecommand \@url [1]{\endgroup\@href {#1}{\urlprefix }}%
\providecommand \urlprefix  [0]{URL }%
\providecommand \Eprint [0]{\href }%
\providecommand \doibase [0]{http://dx.doi.org/}%
\providecommand \selectlanguage [0]{\@gobble}%
\providecommand \bibinfo  [0]{\@secondoftwo}%
\providecommand \bibfield  [0]{\@secondoftwo}%
\providecommand \translation [1]{[#1]}%
\providecommand \BibitemOpen [0]{}%
\providecommand \bibitemStop [0]{}%
\providecommand \bibitemNoStop [0]{.\EOS\space}%
\providecommand \EOS [0]{\spacefactor3000\relax}%
\providecommand \BibitemShut  [1]{\csname bibitem#1\endcsname}%
\let\auto@bib@innerbib\@empty
\bibitem [{\citenamefont {Schaller}\ and\ \citenamefont
  {Klimov}(2004)}]{klimov2004}%
  \BibitemOpen
  \bibfield  {author} {\bibinfo {author} {\bibfnamefont {R.~D.}\ \bibnamefont
  {Schaller}}\ and\ \bibinfo {author} {\bibfnamefont {V.~I.}\ \bibnamefont
  {Klimov}},\ }\href@noop {} {\bibfield  {journal} {\bibinfo  {journal}
  {Letters, Physical Review}\ }\textbf {\bibinfo {volume} {18}},\ \bibinfo
  {pages} {1886601} (\bibinfo {year} {2004})}\BibitemShut {NoStop}%
\bibitem [{\citenamefont {Kamat}(2008)}]{kamat2008}%
  \BibitemOpen
  \bibfield  {author} {\bibinfo {author} {\bibfnamefont {P.~V.}\ \bibnamefont
  {Kamat}},\ }\href {\doibase 10.1021/jp806791s} {\bibfield  {journal}
  {\bibinfo  {journal} {The Journal of Physical Chemistry C}\ }\textbf
  {\bibinfo {volume} {112}},\ \bibinfo {pages} {18737} (\bibinfo {year}
  {2008})}\BibitemShut {NoStop}%
\bibitem [{\citenamefont {Wood}\ and\ \citenamefont
  {Bulovic}(2010)}]{wood2010}%
  \BibitemOpen
  \bibfield  {author} {\bibinfo {author} {\bibfnamefont {V.}~\bibnamefont
  {Wood}}\ and\ \bibinfo {author} {\bibfnamefont {V.}~\bibnamefont {Bulovic}},\
  }\href@noop {} {\bibfield  {journal} {\bibinfo  {journal} {Nano Reviews}\
  }\textbf {\bibinfo {volume} {1}} (\bibinfo {year} {2010})}\BibitemShut
  {NoStop}%
\bibitem [{\citenamefont {Bruchez}\ \emph {et~al.}(1998)\citenamefont
  {Bruchez}, \citenamefont {Moronne}, \citenamefont {Gin}, \citenamefont
  {Weiss},\ and\ \citenamefont {Alivisatos}}]{Bruchez1998}%
  \BibitemOpen
  \bibfield  {author} {\bibinfo {author} {\bibfnamefont {M.}~\bibnamefont
  {Bruchez}}, \bibinfo {author} {\bibfnamefont {M.}~\bibnamefont {Moronne}},
  \bibinfo {author} {\bibfnamefont {P.}~\bibnamefont {Gin}}, \bibinfo {author}
  {\bibfnamefont {S.}~\bibnamefont {Weiss}}, \ and\ \bibinfo {author}
  {\bibfnamefont {A.~P.}\ \bibnamefont {Alivisatos}},\ }\href {\doibase
  10.1126/science.281.5385.2013} {\bibfield  {journal} {\bibinfo  {journal}
  {Science}\ }\textbf {\bibinfo {volume} {281}},\ \bibinfo {pages} {2013}
  (\bibinfo {year} {1998})}\BibitemShut {NoStop}%
\bibitem [{\citenamefont {Michalet}\ \emph {et~al.}(2005)\citenamefont
  {Michalet}, \citenamefont {Pinaud}, \citenamefont {Bentolila}, \citenamefont
  {Tsay}, \citenamefont {Doose}, \citenamefont {Li}, \citenamefont
  {Sundaresan}, \citenamefont {Wu}, \citenamefont {Gambhir},\ and\
  \citenamefont {Weiss}}]{Michalet2005}%
  \BibitemOpen
  \bibfield  {author} {\bibinfo {author} {\bibfnamefont {X.}~\bibnamefont
  {Michalet}}, \bibinfo {author} {\bibfnamefont {F.~F.}\ \bibnamefont
  {Pinaud}}, \bibinfo {author} {\bibfnamefont {L.~A.}\ \bibnamefont
  {Bentolila}}, \bibinfo {author} {\bibfnamefont {J.~M.}\ \bibnamefont {Tsay}},
  \bibinfo {author} {\bibfnamefont {S.}~\bibnamefont {Doose}}, \bibinfo
  {author} {\bibfnamefont {J.~J.}\ \bibnamefont {Li}}, \bibinfo {author}
  {\bibfnamefont {G.}~\bibnamefont {Sundaresan}}, \bibinfo {author}
  {\bibfnamefont {A.~M.}\ \bibnamefont {Wu}}, \bibinfo {author} {\bibfnamefont
  {S.~S.}\ \bibnamefont {Gambhir}}, \ and\ \bibinfo {author} {\bibfnamefont
  {S.}~\bibnamefont {Weiss}},\ }\href {\doibase 10.1126/science.1104274}
  {\bibfield  {journal} {\bibinfo  {journal} {Science}\ }\textbf {\bibinfo
  {volume} {307}},\ \bibinfo {pages} {538} (\bibinfo {year}
  {2005})}\BibitemShut {NoStop}%
\bibitem [{\citenamefont {Norris}\ and\ \citenamefont
  {Bawendi}(1996)}]{norris1996}%
  \BibitemOpen
  \bibfield  {author} {\bibinfo {author} {\bibfnamefont {D.~J.}\ \bibnamefont
  {Norris}}\ and\ \bibinfo {author} {\bibfnamefont {M.~G.}\ \bibnamefont
  {Bawendi}},\ }\href {\doibase 10.1103/PhysRevB.53.16338} {\bibfield
  {journal} {\bibinfo  {journal} {Phys. Rev. B}\ }\textbf {\bibinfo {volume}
  {53}},\ \bibinfo {pages} {16338} (\bibinfo {year} {1996})}\BibitemShut
  {NoStop}%
\bibitem [{\citenamefont {de~Oliveira}\ \emph {et~al.}(1995)\citenamefont
  {de~Oliveira}, \citenamefont {de~Paula}, \citenamefont {Plentz~Filho},
  \citenamefont {Medeiros~Neto}, \citenamefont {Barbosa}, \citenamefont
  {Alves}, \citenamefont {Menezes}, \citenamefont {Rios}, \citenamefont
  {Fragnito}, \citenamefont {Brito~Cruz},\ and\ \citenamefont
  {Cesar}}]{oliveria1995}%
  \BibitemOpen
  \bibfield  {author} {\bibinfo {author} {\bibfnamefont {C.~R.~M.}\
  \bibnamefont {de~Oliveira}}, \bibinfo {author} {\bibfnamefont {A.~M.}\
  \bibnamefont {de~Paula}}, \bibinfo {author} {\bibfnamefont {F.~O.}\
  \bibnamefont {Plentz~Filho}}, \bibinfo {author} {\bibfnamefont {J.~A.}\
  \bibnamefont {Medeiros~Neto}}, \bibinfo {author} {\bibfnamefont {L.~C.}\
  \bibnamefont {Barbosa}}, \bibinfo {author} {\bibfnamefont {O.~L.}\
  \bibnamefont {Alves}}, \bibinfo {author} {\bibfnamefont {E.~A.}\ \bibnamefont
  {Menezes}}, \bibinfo {author} {\bibfnamefont {J.~M.~M.}\ \bibnamefont
  {Rios}}, \bibinfo {author} {\bibfnamefont {H.~L.}\ \bibnamefont {Fragnito}},
  \bibinfo {author} {\bibfnamefont {C.~H.}\ \bibnamefont {Brito~Cruz}}, \ and\
  \bibinfo {author} {\bibfnamefont {C.~L.}\ \bibnamefont {Cesar}},\ }\href
  {\doibase http://dx.doi.org/10.1063/1.114049} {\bibfield  {journal} {\bibinfo
   {journal} {Applied Physics Letters}\ }\textbf {\bibinfo {volume} {66}},\
  \bibinfo {pages} {439} (\bibinfo {year} {1995})}\BibitemShut {NoStop}%
\bibitem [{\citenamefont {Goupalov}\ \emph {et~al.}(2001)\citenamefont
  {Goupalov}, \citenamefont {Suris}, \citenamefont {Lavallard},\ and\
  \citenamefont {Citrin}}]{goupalov2001}%
  \BibitemOpen
  \bibfield  {author} {\bibinfo {author} {\bibfnamefont {S.~V.}\ \bibnamefont
  {Goupalov}}, \bibinfo {author} {\bibfnamefont {R.~A.}\ \bibnamefont {Suris}},
  \bibinfo {author} {\bibfnamefont {P.}~\bibnamefont {Lavallard}}, \ and\
  \bibinfo {author} {\bibfnamefont {D.~S.}\ \bibnamefont {Citrin}},\
  }\href@noop {} {\bibfield  {journal} {\bibinfo  {journal} {Nanotechnology}\
  }\textbf {\bibinfo {volume} {12}},\ \bibinfo {pages} {518} (\bibinfo {year}
  {2001})}\BibitemShut {NoStop}%
\bibitem [{\citenamefont {Neuhauser}\ \emph {et~al.}(2000)\citenamefont
  {Neuhauser}, \citenamefont {Shimizu}, \citenamefont {Woo}, \citenamefont
  {Empedocles},\ and\ \citenamefont {Bawendi}}]{neuhauser2000}%
  \BibitemOpen
  \bibfield  {author} {\bibinfo {author} {\bibfnamefont {R.~G.}\ \bibnamefont
  {Neuhauser}}, \bibinfo {author} {\bibfnamefont {K.~T.}\ \bibnamefont
  {Shimizu}}, \bibinfo {author} {\bibfnamefont {W.~K.}\ \bibnamefont {Woo}},
  \bibinfo {author} {\bibfnamefont {S.~A.}\ \bibnamefont {Empedocles}}, \ and\
  \bibinfo {author} {\bibfnamefont {M.~G.}\ \bibnamefont {Bawendi}},\ }\href
  {\doibase 10.1103/PhysRevLett.85.3301} {\bibfield  {journal} {\bibinfo
  {journal} {Phys. Rev. Lett.}\ }\textbf {\bibinfo {volume} {85}},\ \bibinfo
  {pages} {3301} (\bibinfo {year} {2000})}\BibitemShut {NoStop}%
\bibitem [{\citenamefont {Nirmal}\ \emph {et~al.}(1996)\citenamefont {Nirmal},
  \citenamefont {Dabbousi},\ and\ \citenamefont {Bawendi}}]{nirmal1996}%
  \BibitemOpen
  \bibfield  {author} {\bibinfo {author} {\bibfnamefont {M.}~\bibnamefont
  {Nirmal}}, \bibinfo {author} {\bibfnamefont {B.~O.}\ \bibnamefont
  {Dabbousi}}, \ and\ \bibinfo {author} {\bibnamefont {Bawendi}},\ }\href@noop
  {} {\bibfield  {journal} {\bibinfo  {journal} {Nature}\ }\textbf {\bibinfo
  {volume} {383}},\ \bibinfo {pages} {802} (\bibinfo {year}
  {1996})}\BibitemShut {NoStop}%
\bibitem [{\citenamefont {Frantsuzov}\ \emph {et~al.}(2013)\citenamefont
  {Frantsuzov}, \citenamefont {Volkán-Kacsó},\ and\ \citenamefont
  {Jankó}}]{janko2013}%
  \BibitemOpen
  \bibfield  {author} {\bibinfo {author} {\bibfnamefont {P.~A.}\ \bibnamefont
  {Frantsuzov}}, \bibinfo {author} {\bibfnamefont {S.}~\bibnamefont
  {Volkán-Kacsó}}, \ and\ \bibinfo {author} {\bibfnamefont {B.}~\bibnamefont
  {Jankó}},\ }\href {\doibase 10.1021/nl3035674} {\bibfield  {journal}
  {\bibinfo  {journal} {Nano Letters}\ }\textbf {\bibinfo {volume} {13}},\
  \bibinfo {pages} {402} (\bibinfo {year} {2013})}\BibitemShut {NoStop}%
\bibitem [{\citenamefont {Galland}\ \emph {et~al.}(2011)\citenamefont
  {Galland}, \citenamefont {Ghosh}, \citenamefont {Steinbruck}, \citenamefont
  {Sykora}, \citenamefont {Hollingsworth}, \citenamefont {Klimov},\ and\
  \citenamefont {Htoon}}]{galland2011}%
  \BibitemOpen
  \bibfield  {author} {\bibinfo {author} {\bibfnamefont {C.}~\bibnamefont
  {Galland}}, \bibinfo {author} {\bibfnamefont {Y.}~\bibnamefont {Ghosh}},
  \bibinfo {author} {\bibfnamefont {A.}~\bibnamefont {Steinbruck}}, \bibinfo
  {author} {\bibfnamefont {M.}~\bibnamefont {Sykora}}, \bibinfo {author}
  {\bibfnamefont {J.~A.}\ \bibnamefont {Hollingsworth}}, \bibinfo {author}
  {\bibfnamefont {V.~I.}\ \bibnamefont {Klimov}}, \ and\ \bibinfo {author}
  {\bibfnamefont {H.}~\bibnamefont {Htoon}},\ }\href@noop {} {\bibfield
  {journal} {\bibinfo  {journal} {Nature}\ }\textbf {\bibinfo {volume} {479}},\
  \bibinfo {pages} {203} (\bibinfo {year} {2011})}\BibitemShut {NoStop}%
\bibitem [{\citenamefont {Crouch}\ \emph {et~al.}(2010)\citenamefont {Crouch},
  \citenamefont {Sauter}, \citenamefont {Wu}, \citenamefont {Purcell},
  \citenamefont {Querner}, \citenamefont {Drndic},\ and\ \citenamefont
  {Pelton}}]{crouch2010}%
  \BibitemOpen
  \bibfield  {author} {\bibinfo {author} {\bibfnamefont {C.~H.}\ \bibnamefont
  {Crouch}}, \bibinfo {author} {\bibfnamefont {O.}~\bibnamefont {Sauter}},
  \bibinfo {author} {\bibfnamefont {X.}~\bibnamefont {Wu}}, \bibinfo {author}
  {\bibfnamefont {R.}~\bibnamefont {Purcell}}, \bibinfo {author} {\bibfnamefont
  {C.}~\bibnamefont {Querner}}, \bibinfo {author} {\bibfnamefont
  {M.}~\bibnamefont {Drndic}}, \ and\ \bibinfo {author} {\bibfnamefont
  {M.}~\bibnamefont {Pelton}},\ }\href {\doibase 10.1021/nl100030e} {\bibfield
  {journal} {\bibinfo  {journal} {Nano Letters}\ }\textbf {\bibinfo {volume}
  {10}},\ \bibinfo {pages} {1692} (\bibinfo {year} {2010})}\BibitemShut
  {NoStop}%
\bibitem [{\citenamefont {Ekimov}\ \emph {et~al.}(1993)\citenamefont {Ekimov},
  \citenamefont {Hache}, \citenamefont {Schanne-Klein}, \citenamefont {Ricard},
  \citenamefont {Flytzanis}, \citenamefont {Kudryavtsev}, \citenamefont
  {Yazeva}, \citenamefont {Rodina},\ and\ \citenamefont {Efros}}]{ekimov1993}%
  \BibitemOpen
  \bibfield  {author} {\bibinfo {author} {\bibfnamefont {A.~I.}\ \bibnamefont
  {Ekimov}}, \bibinfo {author} {\bibfnamefont {F.}~\bibnamefont {Hache}},
  \bibinfo {author} {\bibfnamefont {M.~C.}\ \bibnamefont {Schanne-Klein}},
  \bibinfo {author} {\bibfnamefont {D.}~\bibnamefont {Ricard}}, \bibinfo
  {author} {\bibfnamefont {C.}~\bibnamefont {Flytzanis}}, \bibinfo {author}
  {\bibfnamefont {I.~A.}\ \bibnamefont {Kudryavtsev}}, \bibinfo {author}
  {\bibfnamefont {T.~V.}\ \bibnamefont {Yazeva}}, \bibinfo {author}
  {\bibfnamefont {A.~V.}\ \bibnamefont {Rodina}}, \ and\ \bibinfo {author}
  {\bibfnamefont {A.~L.}\ \bibnamefont {Efros}},\ }\href {\doibase
  10.1364/JOSAB.10.000100} {\bibfield  {journal} {\bibinfo  {journal} {J. Opt.
  Soc. Am. B}\ }\textbf {\bibinfo {volume} {10}},\ \bibinfo {pages} {100}
  (\bibinfo {year} {1993})}\BibitemShut {NoStop}%
\bibitem [{\citenamefont {Chepic}\ \emph {et~al.}(1990)\citenamefont {Chepic},
  \citenamefont {Efros}, \citenamefont {Ekimov}, \citenamefont {Ivanov},
  \citenamefont {Kharchenko}, \citenamefont {Kudriavtsev},\ and\ \citenamefont
  {Yazeva}}]{chepic1990}%
  \BibitemOpen
  \bibfield  {author} {\bibinfo {author} {\bibfnamefont {D.~I.}\ \bibnamefont
  {Chepic}}, \bibinfo {author} {\bibfnamefont {A.}~\bibnamefont {Efros}},
  \bibinfo {author} {\bibfnamefont {A.~I.}\ \bibnamefont {Ekimov}}, \bibinfo
  {author} {\bibfnamefont {M.~G.}\ \bibnamefont {Ivanov}}, \bibinfo {author}
  {\bibfnamefont {V.~A.}\ \bibnamefont {Kharchenko}}, \bibinfo {author}
  {\bibfnamefont {I.~A.}\ \bibnamefont {Kudriavtsev}}, \ and\ \bibinfo {author}
  {\bibfnamefont {T.~V.}\ \bibnamefont {Yazeva}},\ }\href {\doibase
  http://dx.doi.org/10.1016/0022-2313(90)90007-X} {\bibfield  {journal}
  {\bibinfo  {journal} {Journal of Luminescence}\ }\textbf {\bibinfo {volume}
  {47}},\ \bibinfo {pages} {113 } (\bibinfo {year} {1990})}\BibitemShut
  {NoStop}%
\bibitem [{\citenamefont {Efros}\ and\ \citenamefont
  {Rosen}(2000)}]{efros2000}%
  \BibitemOpen
  \bibfield  {author} {\bibinfo {author} {\bibfnamefont {A.~L.}\ \bibnamefont
  {Efros}}\ and\ \bibinfo {author} {\bibfnamefont {M.}~\bibnamefont {Rosen}},\
  }\href {\doibase 10.1146/annurev.matsci.30.1.475} {\bibfield  {journal}
  {\bibinfo  {journal} {Annual Review of Materials Science}\ }\textbf {\bibinfo
  {volume} {30}},\ \bibinfo {pages} {475} (\bibinfo {year} {2000})}\BibitemShut
  {NoStop}%
\bibitem [{\citenamefont {Dabbousi}\ \emph {et~al.}(1997)\citenamefont
  {Dabbousi}, \citenamefont {Rodriguez-Viejo}, \citenamefont {Mikulec},
  \citenamefont {Heine}, \citenamefont {Mattoussi}, \citenamefont {Ober},
  \citenamefont {Jensen},\ and\ \citenamefont {Bawendi}}]{Dabbousi1997}%
  \BibitemOpen
  \bibfield  {author} {\bibinfo {author} {\bibfnamefont {B.~O.}\ \bibnamefont
  {Dabbousi}}, \bibinfo {author} {\bibfnamefont {J.}~\bibnamefont
  {Rodriguez-Viejo}}, \bibinfo {author} {\bibfnamefont {F.~V.}\ \bibnamefont
  {Mikulec}}, \bibinfo {author} {\bibfnamefont {J.~R.}\ \bibnamefont {Heine}},
  \bibinfo {author} {\bibfnamefont {H.}~\bibnamefont {Mattoussi}}, \bibinfo
  {author} {\bibfnamefont {R.}~\bibnamefont {Ober}}, \bibinfo {author}
  {\bibfnamefont {K.~F.}\ \bibnamefont {Jensen}}, \ and\ \bibinfo {author}
  {\bibfnamefont {M.~G.}\ \bibnamefont {Bawendi}},\ }\href@noop {} {\bibfield
  {journal} {\bibinfo  {journal} {The Journal of Physical Chemistry B}\
  }\textbf {\bibinfo {volume} {101}},\ \bibinfo {pages} {9463} (\bibinfo {year}
  {1997})}\BibitemShut {NoStop}%
\bibitem [{\citenamefont {Empedocles}\ and\ \citenamefont
  {Bawendi}(1997)}]{empedocles1997}%
  \BibitemOpen
  \bibfield  {author} {\bibinfo {author} {\bibfnamefont {S.~A.}\ \bibnamefont
  {Empedocles}}\ and\ \bibinfo {author} {\bibfnamefont {M.~G.}\ \bibnamefont
  {Bawendi}},\ }\href {\doibase 10.1126/science.278.5346.2114} {\bibfield
  {journal} {\bibinfo  {journal} {Science}\ }\textbf {\bibinfo {volume}
  {278}},\ \bibinfo {pages} {2114} (\bibinfo {year} {1997})}\BibitemShut
  {NoStop}%
\bibitem [{\citenamefont {Empedocles}\ and\ \citenamefont
  {Bawendi}(1999)}]{emp1999}%
  \BibitemOpen
  \bibfield  {author} {\bibinfo {author} {\bibfnamefont {S.~A.}\ \bibnamefont
  {Empedocles}}\ and\ \bibinfo {author} {\bibfnamefont {M.~G.}\ \bibnamefont
  {Bawendi}},\ }\href@noop {} {\bibfield  {journal} {\bibinfo  {journal} {The
  Journal of Physical Chemistry B}\ }\textbf {\bibinfo {volume} {103}},\
  \bibinfo {pages} {1826} (\bibinfo {year} {1999})}\BibitemShut {NoStop}%
\end{thebibliography}%

\end{document}